# Localization of light in a random grating array in a single mode fiber

Ofer Shapira and Baruch Fischer

*Department of Electrical Engineering, Technion - Israel Institute of Technology,*

*Haifa 32000, Israel*

*E-mail: Ofer Shapira: ofers@MIT.EDU , Baruch Fischer: fischer@ee.technion.ac.il*

## Abstract

We investigate light propagation in randomly spaced fiber gratings in a single mode fiber, and demonstrate the localization effect. Localization of light in random media resembles that of electrons in disordered solids, resulting from a subtle wave interference formation. We measured the light transmission after each additional grating fabrication and found an exponential decay that follows the localization theory. An important feature of the random array is its similarity to ordered gratings in the transmission and reflection behavior at the long array regime. Besides the basic interest in localization in one-dimensional systems, random grating arrays have potential applications, utilizing the possibility to fabricate long structures with strong and broadband reflections.

## 1. Introduction

Wave propagation in random media has been an important research topic throughout the years, gaining much attention when the concept of localization appeared and stimulated a large amount of work. The idea of localization was first raised by Anderson [1] for electrons in disordered solids which were drastically affected by quantum mechanical wave interference [2-3]. The quest to study and realize such effects in optics was natural, and indeed a considerable amount of work can be found on light propagation in random media that include aspects of localization. Using light, in lieu of electrons, for the study of localization adds new possibilities, mainly in the experimental aspects due to the relatively easy measuring techniques and the direct access to the optical "wave function" via light intensity measurement.

Properties of wave propagation in random media, including localization, depend in general on the system dimensionality. The theoretical analysis of one dimensional system is obviously easier, than for higher dimensions, but is not at all trivial for experimental realizations in solid state physics. In optics, however, the experimental situation is very different, and 1D wave propagation is trivial. There were many works on localization aspects with electromagnetic waves in the optical [5-15] as well as the microwave [16] regimes. We point out a work by Berry and Klein [5] that is very relevant to our present study. It was shown in a simple but remarkable experiment that for a stack of *N* transparent plates with randomly varying thickness the transmitted intensity decays exponentially with *N*. The striking feature of the random optical elements is that their overall transmissivity $\tau_N$ is given by a simple multiplication of the single element transmissivity $\tau$, i.e. $\tau_N = \tau^N$. It means that only the direct transmission counts, while any multiple reflections added in the direction of the transmitted light interfere destructively. We also presented in previous work two experimental realizations for localization of light in optical kicked-rotor, which resemble the quantum kicked-rotor that relates to Anderson localization [11-15]. In the first [11-12], we demonstrated localization in the spatial frequency domain of free space light beam propagating through an array of thin sinusoidal phase gratings. In the second case [13-15], we studied the spectrum (or sidebands) localization of light pulses which are repeatedly "kicked" by a sinusoidal RF modulation along a fiber.

In this paper we present an experimental study of 1D localization by means of light propagation in a random Bragg grating array fabricated into a single mode fiber. We demonstrate the localization behavior, manifested in the strong exponential decay of the light transmission along the fiber, that was measured directly after the fabrication of each additional grating. This decay, which results in high reflection, should not be confused with the much smaller fiber loss. A first report on this finding was given in Ref. [19]. The theoretical analysis of such system is based on the transfer matrix formalism in which the system is represented by a product of random matrices. The asymptotic behavior of such a product results from a theorem on products of random matrices by Furstenberg [18]. This theorem ensures that under very general conditions, the elements of the matrix product and any norm of the matrix product grow exponentially with the same exponent, giving rise to the localization behavior. We refer the reader to a comprehensive analysis for 1D disordered system given by Pendry [3]. Beside the basic propagation effects in the random array, the study can have importance ramifications on fiber-optic communication and gratings technology. Examples are strong and broadband reflectors, fiber lasers and random lasers.

The scattering elements in our system are the random gratings in single-mode fibers. Fibers are an ideal experimental medium for 1D light propagation with a matured technology of in-fiber grating fabrication. The gratings were made to be almost identical, with the same and relatively large reflection wavelength bandwidth of a few *nm*, obtained by fabricating short gratings. Therefore the interesting effects concerning the localization occur within that bandwidth. Gratings are very effective scattering elements such that we could observe the localization effect with a relatively small number of them, about 50 gratings. The randomness of the scattering elements enters by the random spacing between the gratings (see Fig. 1.1).

The outline of the paper chapters is as follows: We first give in chapter 2 theoretical treatment of the wave propagation in randomly spaced gratings. We then compare the wave theory result for the light transmissivity to the calculation obtained from the ray theory and also to wave propagation in an ordered fiber grating. Chapter 3 describes the experiment, starting with the setup and the grating fabrication system, and then presenting the experimental results. We show transmission measurements, the spectra and the transmissivity as a function of the gratings number for the random fiber system. These curves are the central results of the paper, showing the localization effect in the random grating array. We then compare the measured results to the theory and find a very good agreement. We end the paper with conclusions and remarks on the application sides.

# 2. Theoretical Analysis of Light Transmissivity in a 1D Random Grating Array

We present a theoretical treatment for wave propagation in a single-mode fiber with *N* randomly-spaced Bragg gratings and calculate the transmission in the limit of *N>>1*, to obtain the localization length. This result readily reveals that the interference among all reflected waves is destructive for the transmission and intuitive explanation is presented. We then compare this analysis to *ray theory,* in which light is treated as lacking phase property, to show that contrary to the exact wave calculation, this theory results in transmission that decays as *1/N.* Finally we calculate the transmission of ordered system and compare the decay rates in both cases.

## 2.1 Disordered Grating Array

Transmission calculation through a disordered chain of gratings can be carried out by transfer matrices methods [17]. The basic idea underlying such calculation assumes that the system can by cut into slices, when each can be easily evaluated. Then, by writing the transfer matrix of the complete system as a product of those matrices, we can apply Furstenberg theorem [18] to obtain the asymptotic behavior of the product.

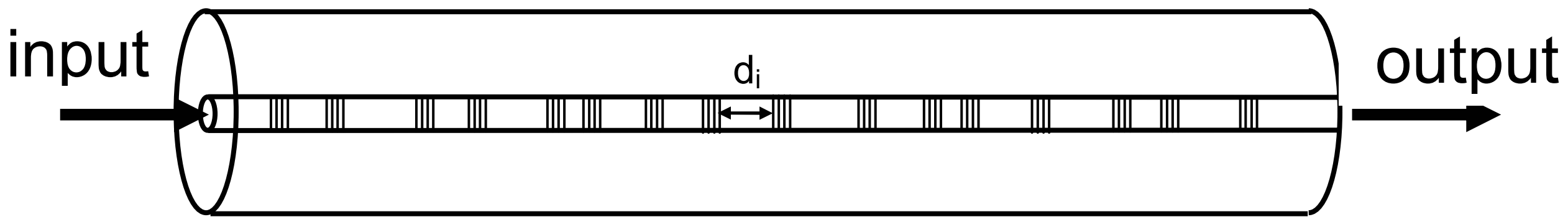

Figure 1.1: Random fiber grating array

The grating system is described in Fig. 1.1. Light with wave-number $k$ propagates along a single mode fiber having an array of successive randomly-spaced gratings. It will be assumed that the space widths, $d_i$, are drawn independently from the density distribution function $d\mu(d_i)$, and that the gratings are identical i.e., have the same lengths and refractive indices.

We describe the propagation of the light in the 1D medium by the transfer matrix, $M_i$ that relates the amplitudes of the forward and backward propagating waves on the right of each optical element to those on the left:

$$(2.1)\quad \begin{bmatrix} a_n^{+} \\ a_n^{-} \end{bmatrix} = M_n \begin{bmatrix} a_{n-1}^{+} \\ a_{n-1}^{-} \end{bmatrix}$$

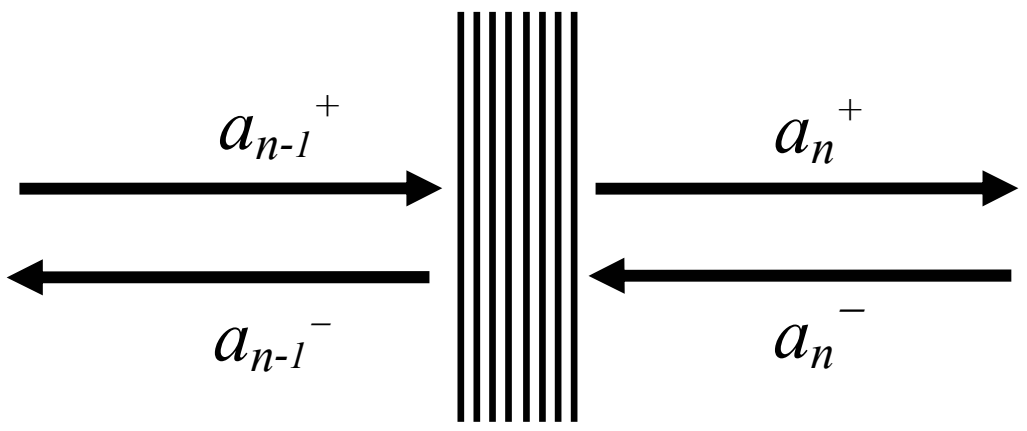

Figure 1.2: Incident and reflected field amplitudes that define the transfer matrix of a single grating.

For optical lossless elements that are invariant under time reversal, the scattering matrix (that relates the amplitudes of the ingoing to the outgoing waves, via reflections and transmissions) is unitary. Then, by denoting the amplitude reflection and transmission coefficients from both sides, for incidence from the left $\boldsymbol{r_n}$, , $\boldsymbol{t_n}$ and from the right $\boldsymbol{r'_n}$and $\boldsymbol{t'_n}$, we have the relations: $\boldsymbol{r'_n t_n^* + r_n^* t'^*_n = 0}$ , and $| \boldsymbol{r_n} |^2 + | \boldsymbol{t_n} |^2 = | \boldsymbol{r'_n} |^2 + | \boldsymbol{t'_n} |^2 = \boldsymbol{1}$ , ( the asterisk * stands for complex conjugate) and the transfer matrix is given by [5, 20]:

$$(2.1)\quad m_n = \begin{pmatrix} \frac{1}{t_n^*} & -\frac{r_n^*}{t_n^*} \\ -\frac{r_i}{t_n} & \frac{1}{t_n} \end{pmatrix}.$$

The transfer matrices are unimodular (det $\mathbf{m_n}$=1). In our system we define each element as being comprised of one grating and its successive space. The gratings are taken to be identical, and the spacing between are responsible for the random part. The transfer matrix of such element is the product of the grating transfer matrix and the space transfer matrix. From coupled wave equation of the counter-propagating waves, the transfer matrix for a single grating is given by [21]:

$$(2.2)\quad m_g = \begin{pmatrix} \cosh(SL) - i\frac{\Delta\beta}{S}\sinh(SL_0) & -i\frac{\kappa}{S}\sinh(SL_0) \\ i\frac{\kappa}{S}\sinh(SL_0) & \cosh(SL) + i\frac{\Delta\beta}{S}\sinh(SL_0) \end{pmatrix}$$

where $L$ is the single grating length, κ is the coupling coefficient between the counter-propagating beams in the gratings, $\Delta\beta = \beta - \pi / \Lambda$ is the wave-number deviation from the Bragg wavelength, Λ is the grating period, and $S = \sqrt{|\kappa|^2 - \Delta\beta^2}$ .

The transfer matrix for a space of length $d_i$ is given by

$$(2.3)\quad m_{d_i} = \begin{pmatrix} \exp(ikd_i) & 0 \\ 0 & \exp(-ikd_i) \end{pmatrix}$$

Then the transfer matrix for a single element is the product of the two above matrices:

$$(2.4)\quad m_i = m_g m_{d_i} = \begin{pmatrix} \left[\cosh(SL) - i\frac{\Delta\beta}{S}\sinh(SL_0)\right]\exp(ikd_i) & -i\frac{\kappa}{S}\sinh(SL_0)\exp(-ikd_i) \\ i\frac{\kappa}{S}\sinh(SL_0)\exp(ikd_i) & \left[\cosh(SL) + i\frac{\Delta\beta}{S}\sinh(SL_0)\right]\exp(-ikd_i) \end{pmatrix}$$

and the single element transmission and reflection coefficients are given by:

$$(2..5)\quad t_i = [m_i]_{22} = \left[\cosh(SL_0) - i\frac{\Delta\beta}{S}\sinh(SL_0)\right]^{-1}\exp(ikd_i)$$

(2..6) $r_i = -[m_i]_{11} / [m_i]_{22} = \kappa \exp(ikd_i) \sinh(SL_0) / [iS\cosh(SL_0) - \Delta\beta \sinh(SL_0)]$

$\alpha_i = kd_i$ provides the random nature of the system when we have a set of such elements. It is assumed that the space widths $d_i$, are drawn independently from a density distribution function $d\mu(\alpha_i)$. The transfer matrix for $N$ gratings and $N$ spaces is:

$$(2.7)\quad M_N = m_1 m_2 ....m_N = \begin{pmatrix} \frac{1}{T_N^*} & -\frac{R_N^*}{T_N^*} \\ -\frac{R_N}{T_N} & \frac{1}{T_N} \end{pmatrix}$$

$T_N$ and $R_N$ are the amplitude transmission and reflection coefficients, respectively, for the entire system. All coefficients as well as the transfer matrix of the complete optical array are denoted in this paper by capital letters: T, R, M, compared to t, r, m, for one element. For the use below we also denote the intensity transmissivity and reflectivity for a single grating by: $\tau = |t|^2$, $t = \tau \exp(\theta)$ and $\rho = |r|^2$, and for the array of N gratings: $\tau_N = |T_N|^2$ and $\rho_N = |R_N|^2$.

We next evaluate the product of those $N$ random unimodular matrices of Eq. 2.7, to obtain the system overall transmission. The asymptotic behavior of $\boldsymbol{M_N}$ can be obtained using Furstenberg theorem [18] on the product of random matrices, stating that under very general conditions, the elements of the matrix product and any norm of the matrix product grow exponentially with the same exponent:

$$(2.8)\quad \frac{1}{N}\log\|m_N....m_1 u\| \rightarrow \iint \log\left|\frac{m(\alpha)u}{u}\right| d\mu(\alpha) d\nu(\theta) \equiv \gamma$$

where,

$$(2.9)\quad \nu(\theta) = \int \nu(\theta(\alpha)) \frac{d\theta(\alpha)}{d\theta} d\mu(\alpha)$$

defines the probability distribution of u, and θ=arg(u). For the transfer matrix given at (2.1), and in fact for more general case, Furstenbergs conditions are satisfied as shown by Matsuda and Ishii [4]. Then from (2.8) and (2.9) the exponent is given by

$$(2.10)\quad \lim_{N\to\infty}\frac{1}{N}\ln\tau_N = -\ln(1/\tau)$$

where τ is the transmission of a single grating, and the system overall transmission is

$$(2.11)\quad \tau_N = e^{-N\ln(1/\tau)} = \tau^N\ .$$

This simple result reveals an interesting property of the transmission through a set of randomly spaced scatterers in a 1D system by showing that it comprises only from the multiplication of the single gratings transmission and doesn't include all multiple reflections. Of course, reflections were taken into consideration in the above calculation but asymptotically the result teaches us that all multiple reflections are canceled.

## 2.2 Numerical Simulation for Disordered Grating Array

In the previous section we obtained the transmissivity for a large number of disordered gratings and found it to be exponentially decaying with the number of gratings. Here we compare this analytical result to a numerical simulation of the transmission. Figures 2.1 and 2.2 depict the transmitted intensity spectrum after 1,000 and 5,000 gratings, respectively. The gratings were taken to be identical, and have the following properties: centered at 1540*nm*, their coupling coefficient κ = 345$m^{-1}$, and their length L = 0.385*mm*. The transmissivity for a single grating at the band center is 0.0483*dB*. The distances between two successive gratings were chosen randomly from the interval [0–1] *mm*. The figures also present the transmission spectrum given by Equation 2.11, with the wavelength dependent transmission $\tau$, of a single grating having the same parameters as those given above. In both cases, a very good agreement was obtained between the analytical calculation and the numerical simulation. The smoother nature of the longer array is obvious, as the averaging action over many gratings is more uniformly spread. One can view the array output spectrum as being composed from all grating pairs making many random Fabry-Perot etalons. The output is the collective spectra which gradually lose the individual Fabry-Perot characteristic as the light passes more gratings. Fig. 2.3 shows the evolution of the transmitted intensity at the band center after each grating. Here, too, the good agreement between the analytical calculation and the numerical simulation is well observed.

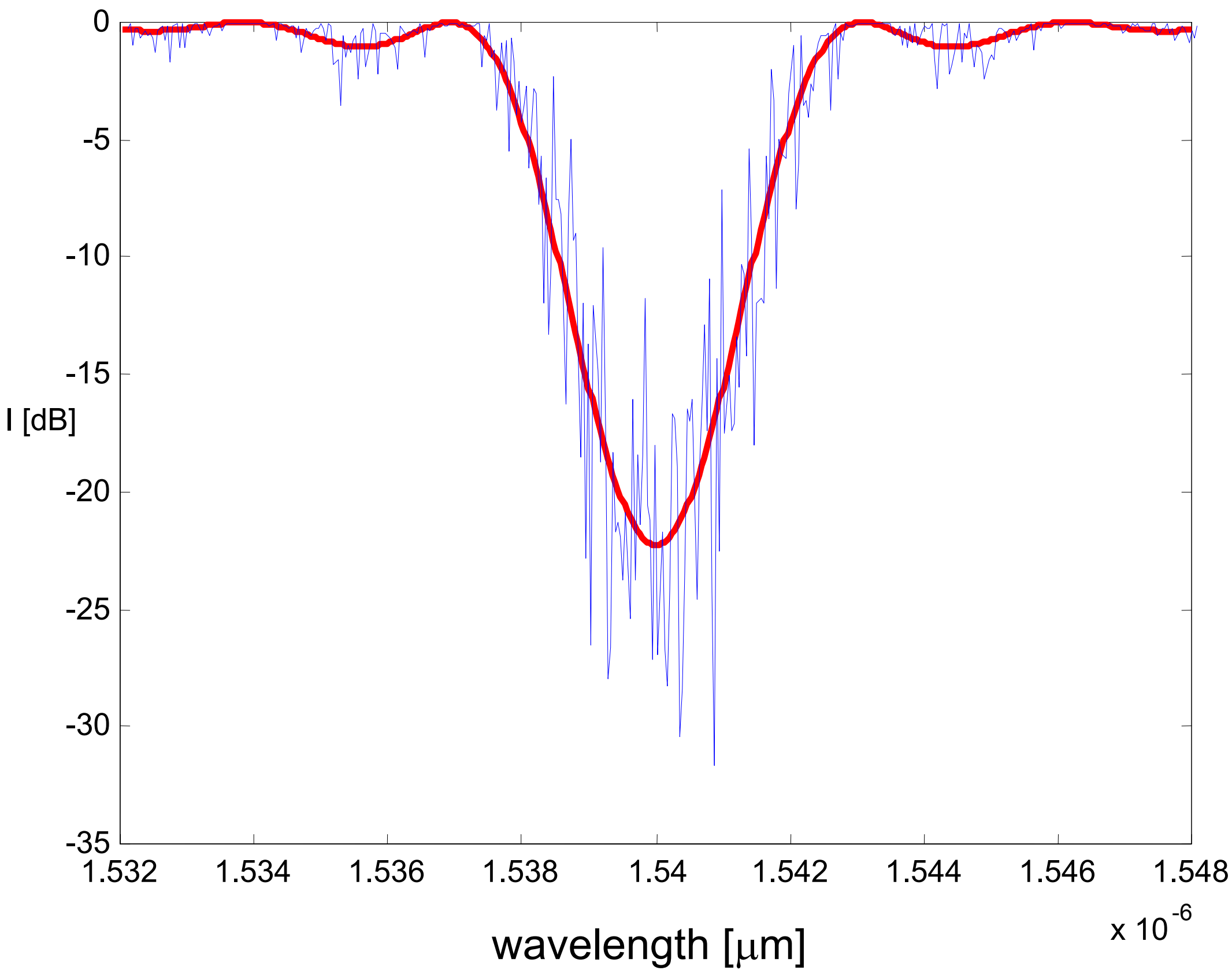

Figure 2.1: Transmission spectrum simulation of 1,000 randomly-spaced gratings with a single grating transmissivity of 0.0483*dB* at the band center. The continuous line shows the asymptotic behavior of the transmission spectrum.

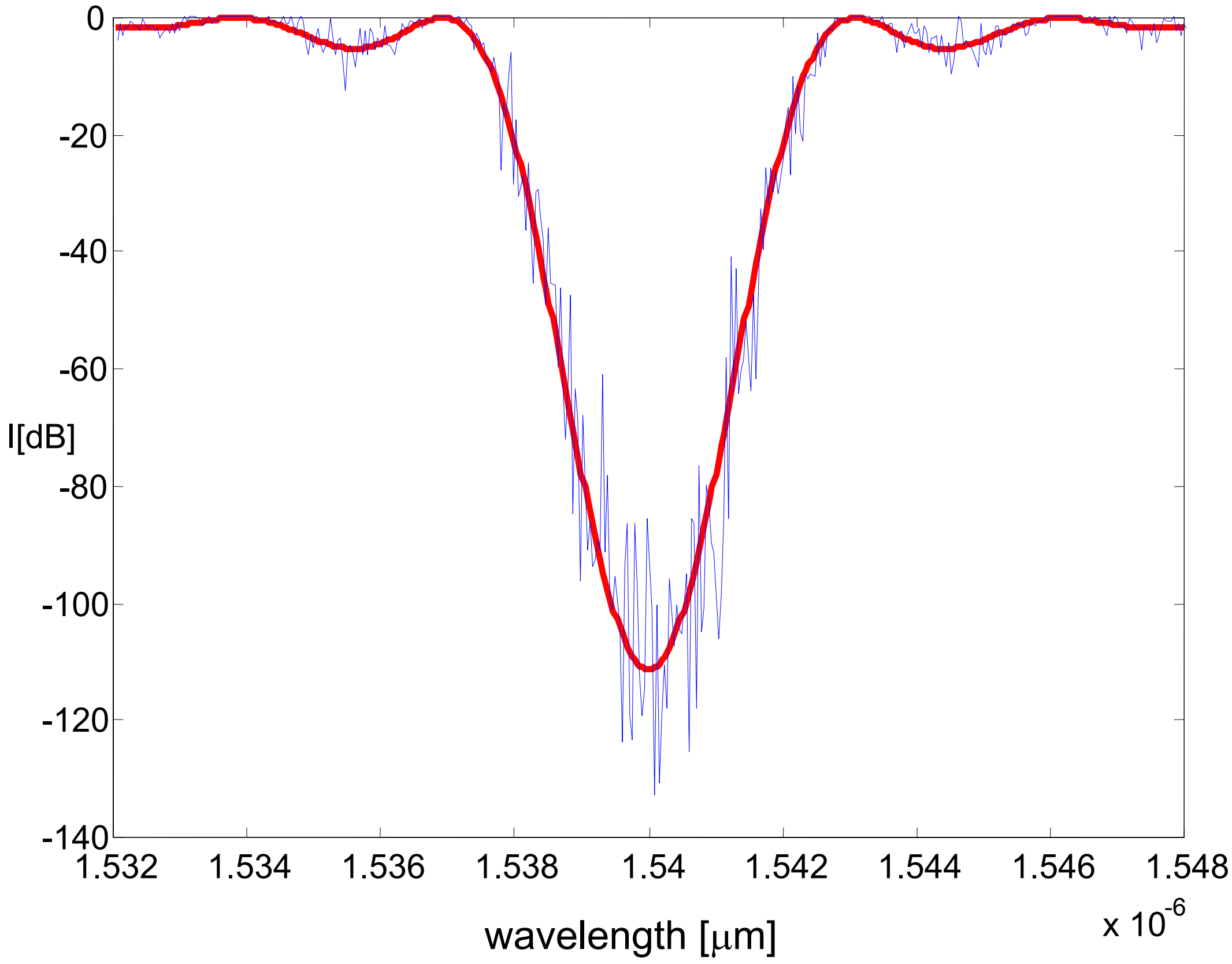

Figure 2.2: Transmission spectrum simulation of 5,000 randomly-spaced gratings with a single grating transmissivity of 0.0483*dB* at the band center. The continuous line shows the asymptotic behavior of the transmission spectrum.

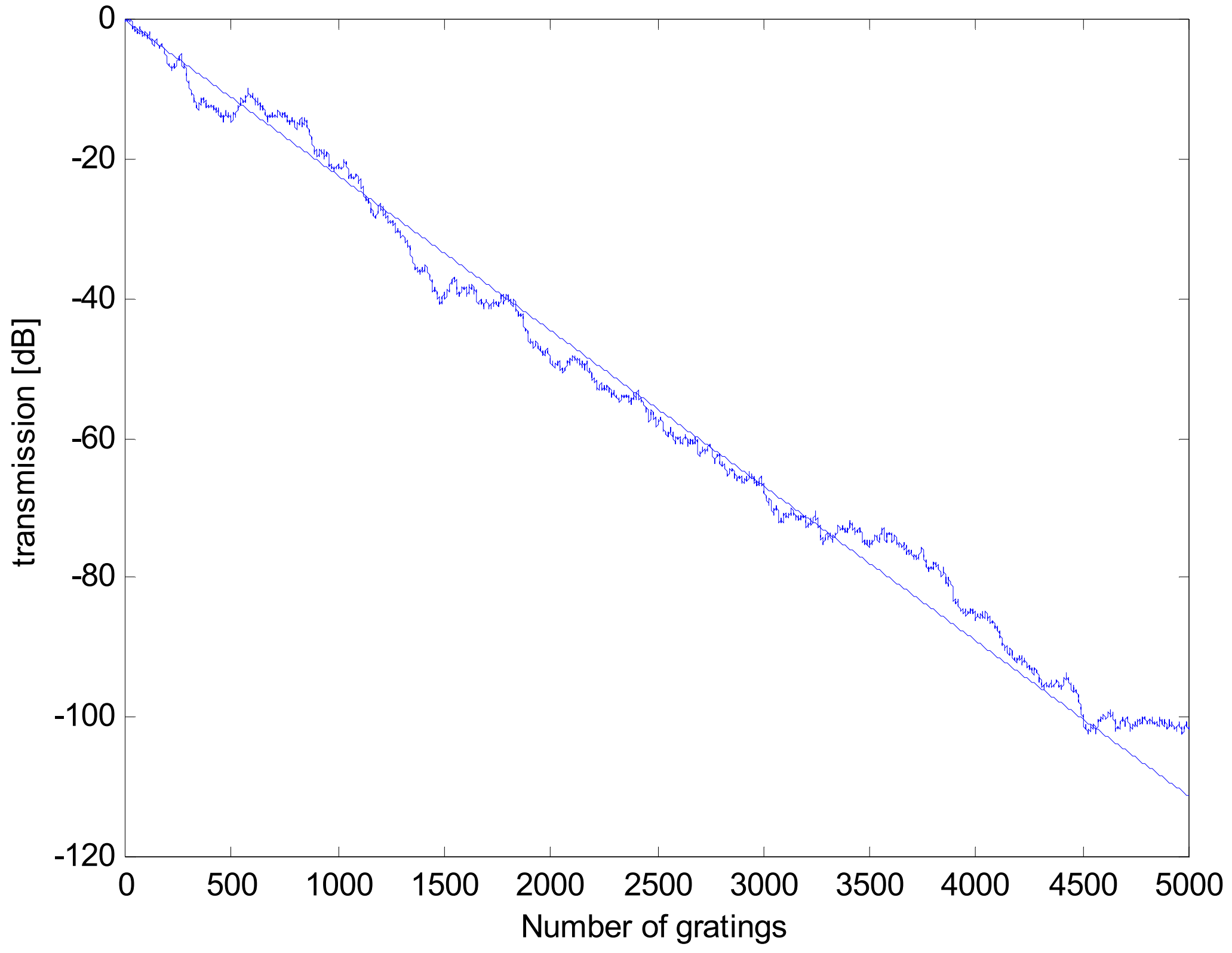

Figure 2.3: Transmissivity at the band center after each grating: Simulation and the asymptotic behavior (the straight line) given by the theory for a single grating transmissivity of 0.0483$dB$.

### 2.3 Destructive Interference of the High Order Reflections

It is possible to consider the total transmission as an infinite sum of waves formed by multiple reflections and transmissions consisting of different optical paths, and different phases. Fig. 2.4 is an example of a system built out of six randomly-spaced gratings. The figure exemplifies three waves with exactly the same overall path length, but with different number of transmissions and reflections. Nevertheless, due to the phase difference between the reflection coefficient of the forward and backward propagating waves, the two upper waves interfere constructively, however the third wave interferes destructively with the upper two. The fascinating result of the localization theory is that for large arrays the overall interference of the all multi-reflections in the transmitted light is fully destructive. Therefore, the total transmission is comprised only of the wave that passes through all elements without being reflected.

The phase difference between the reflection coefficients for opposite wave incidence at an optical element is general. We are familiar with the opposite sign of the reflection coefficients for opposite incident waves at a boundary of two media with different refractive indices. More generally it can be tracked in the relation mentioned in section 2.1: $r'_n t_n^* + r_n^* t'^*_n = 0$. For these optical elements $|r_n|=|r'_n|$, $t_n= t'_n$ and for specific choice of the reference planes of the waves at the two sides of the element, $\arg(t_n) = \arg(t'_n) = 0$, and then obtain $r_n = -r'_n$.

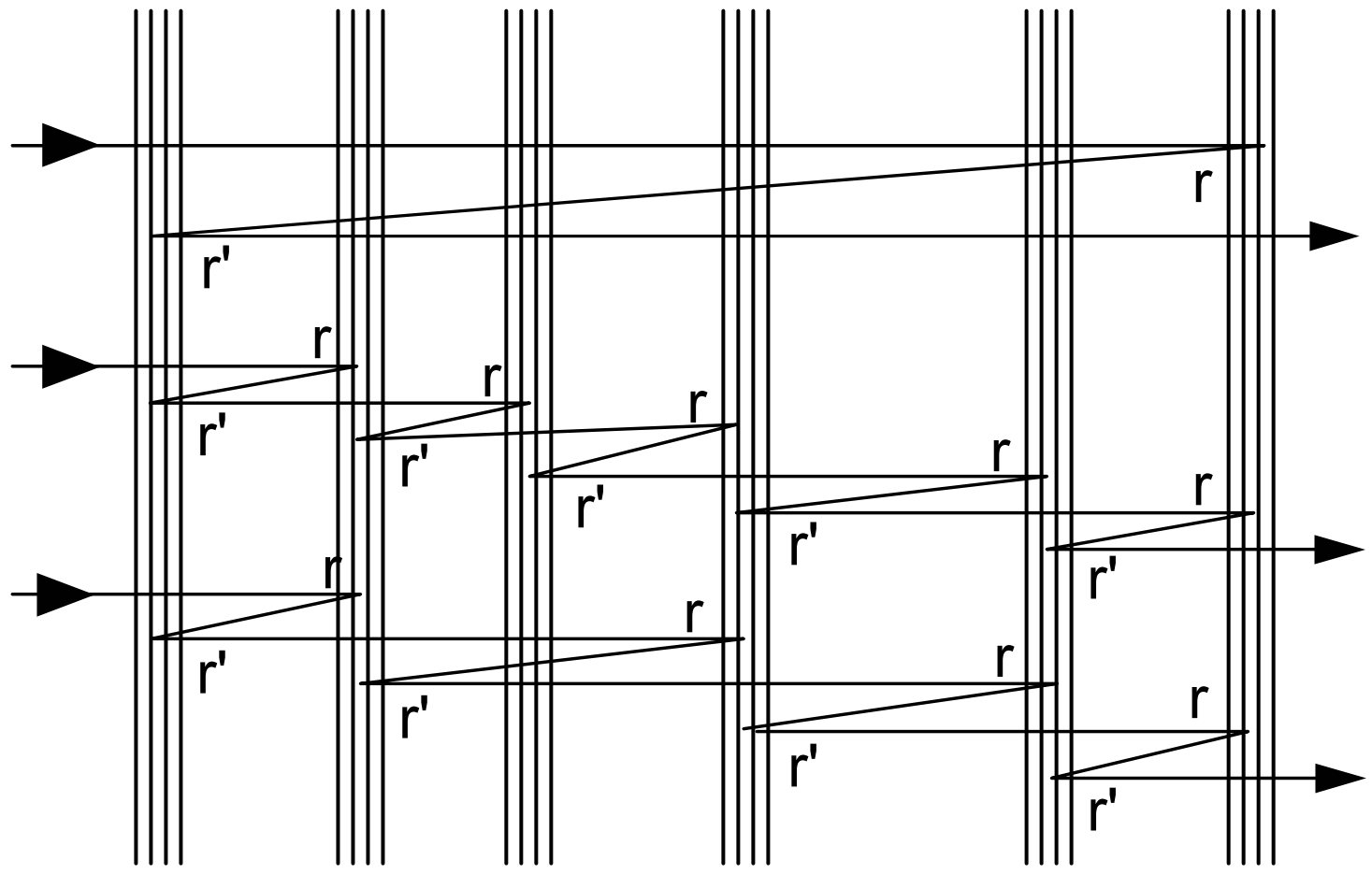

Figure 2.4: Three waves with paths of the same length, but different numbers of reflections resulting in constructive interference between the two upper waves but destructive interference with the third.

### 2.4 Ray Theory for Disordered System

We show here the simple ray theory approach, that could have been naively expected to be adequate for the disordered system, but in fact it leads to wrong results. The development follows the work by Berry and Klein [5] given here to clarify the basic wave nature responsible for the localization effect. Ray theory approach is based upon the assumption that the waves in a disordered system are incoherent; and therefore can be represented as intensities rather than amplitudes. The appropriate matrix formalism can be obtained for the ray theory, referring to incident and reflected intensities. When $\rho$ and $\tau$ are the one element intensity reflectivity and transmissivity, where for lossless scatterers $\tau + \rho = 1$, the one element transfer matrix is:

$$(2.12)\quad m = \begin{pmatrix} \tau - \dfrac{\rho^2}{\tau} & \dfrac{\rho}{\tau} \\ -\dfrac{\rho}{\tau} & \dfrac{1}{\tau} \end{pmatrix}$$

and for N successive elements,

$$(2.13)\quad m^N = \begin{pmatrix} \tau - \dfrac{\rho^2}{\tau} & \dfrac{\rho}{\tau} \\ -\dfrac{\rho}{\tau} & \dfrac{1}{\tau} \end{pmatrix}^N = \left[ \mathrm{I} + \frac{\rho}{\tau} \begin{pmatrix} -1 & 1 \\ -1 & 1 \end{pmatrix} \right]^N = \mathrm{I} + N\frac{\rho}{\tau} \begin{pmatrix} -1 & 1 \\ -1 & 1 \end{pmatrix}$$

I is the unit matrix, and the last equality is based on $\begin{pmatrix} -1 & 1 \\ -1 & 1 \end{pmatrix}^2 = 0$ . Therefore, the ray theory transmissivity for $N$ random gratings is:

$$(2.14)\quad \tau_N = m_{22}{}^{-1} = \frac{\tau}{\tau + N(1-\tau)}$$

It is a linear decay, or Ohmic like behavior for $(1/T_N) \propto N$, (for large $N$), which is fundamentally different from the exponential dependence results from localization theory. Fig. 2.5 shows graphically the transmissivity in the two approaches.

The different and misunderstood result of the ray theory lays in the fact that regarding waves as incoherent and averaging over random phases is not equivalent. This difference is exemplified in section 2.3: although propagating in a random medium, different light wave paths of the same lengths "magically" give precise destructive interferences. Thus, the assumption that the scattered waves have no phase correlation, and therefore can be regarded as incoherent, is false. Furthermore, exact wave averaging shows that all transmitted waves (except for the one passing without any reflection) interfere destructively, leading to the exponential decay of the transmissivity. It is also noteworthy that the ray theory does give a correct result when the waves interference is not dominant. This can occur when the reflection is very small and the system is small enough so even small reflections would not accumulate.

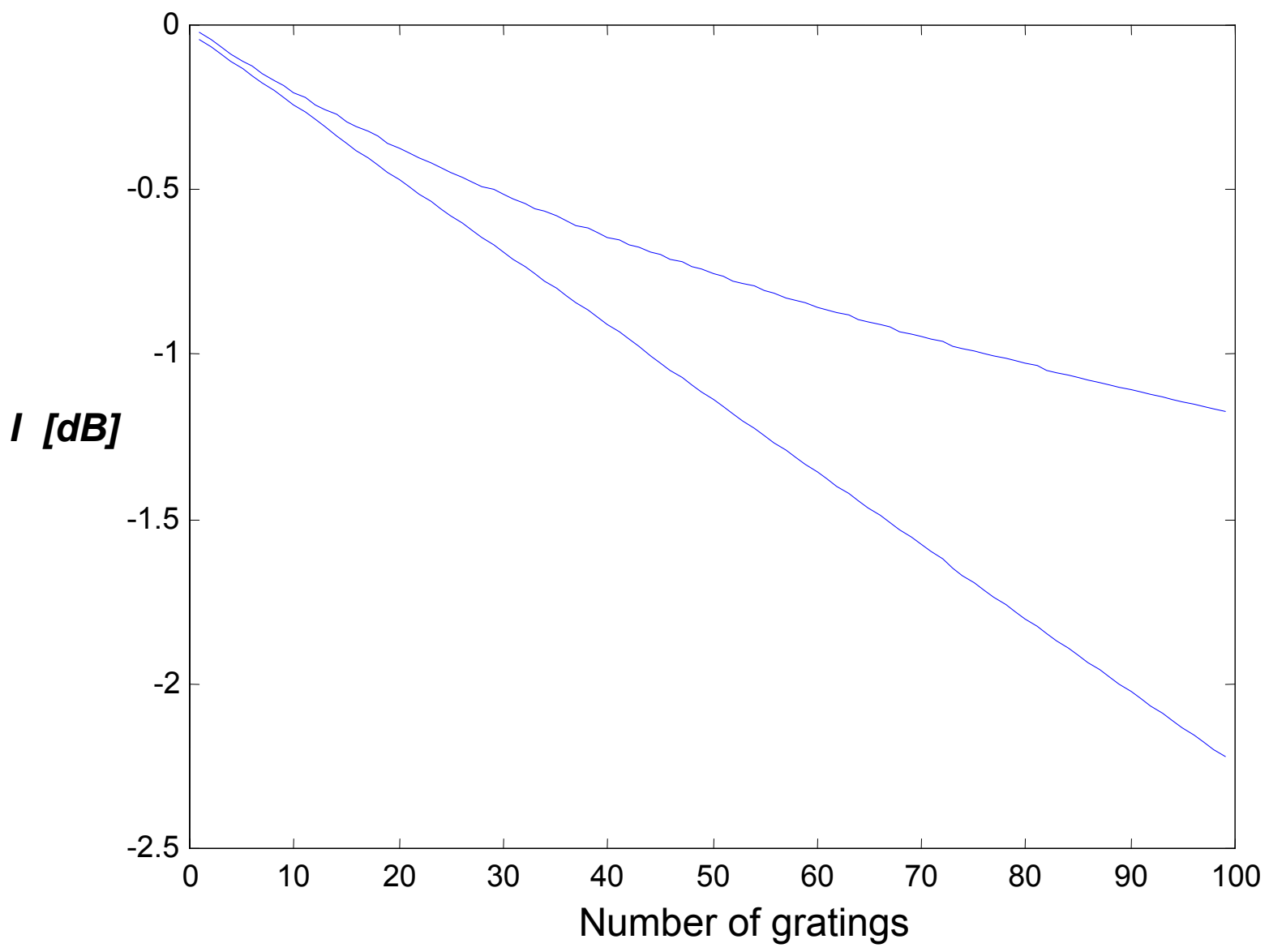

Figure 2.5: Comaprison between wave and ray theories for the transmissivity in random gratings: the upper line depicts the Ohmic behavior of the ray theory while the lower depicts the exact wave averaging.

## 2.5 Comparison to Ordered Gratings

It is interesting to compare the random grating system to ordered grating. It seems that we have a powerful method to obtain effectively long gratings, winning with the easy fabrication of random structures that might have been regarded as a spoiling factor, but turns out to be an advantage with the support of the localization effect that provides a strong transmission decay and high reflection. Then, not only that the random grating arrays are easy to fabricate, but disturbances don't have much deteriorating effect on their performance. On the other hand, long ordered gratings are hard to fabricate and can have detrimental environment effect. The situation is even worse for ordered grating arrays, which are almost impossible to implement even for a small number of gratings, since they need precise interferometric spacing between the gratings. An additional advantage of the random arrays is that they can easily provide very large bandwidths for the reflection, since they depend on the single grating bandwidth. The single grating can be made very short (10-100 μm long) and still be an effective scatterer, thus providing very large wavelength bandwidths of tens of nanometers. One can then argue that for even larger bandwidths, why not to take it to the ultimate situation by forming point scatterers rather then short gratings. Such random point scatterer arrays are of course interesting and worth for implementation, though it is not easy to make. It needs the fabrication of many random scatterers along the fiber. For reaching the asymptotic noiseless localization regime, we need a reasonable number of scatteres. It means that the strength of one scatterer ought to be weak to allow the laser to penetrate and acquire many multi-reflections which averages to zero in the transmitted light. On the other hand, the fabrication of many scatterers is more complicated and therefore we need a reasonable scattering strength for each element to be able to observe the effect. We also mention some disadvantages of the random array. The transmission and the reflection and their spectra aren't smooth and uniform, as the averaging in a random structure which is limited in length, is not optimal. We need a rather long array to reach the smoother asymptotic behavior. The last point in the ordered-disordered gratings comparison is that gratings are mostly used for precise filtering purposes and not only for reflection, and thus the ordered element is needed, unless we look for special filtering uses, or with special finger prints.

For a comparison between the disordered array and ordered structures, like long uniform gratings, we take them to be of the same overall length $L=NL_0$. We can also extend the comparison to ordered arrays of *N* gratings, each of length $L_0$, with exactly the same spacing between them. We note that the latter structure is almost impossible to implement even for low number of gratings, because of the sub-wavelength spacing requirement.

We use Eq. 2.2 for the transmisivity of long uniform grating, replacing $L_0$ by *L*, and then compare the outcome to the random grating result. For the ordered grating array we can again start with the transfer matrix formalism, using Eqs. 2.4 and 2.7, and setting equal spacing, $d_i = d$. This structure include the ordered grating case for *d=0*. Then the transfer matrix for a set of N gratings is:

(2.15) $M_N = m^N$

$M_N$ is given for a unimodular matrix by:

(2.16) $M_N = \begin{pmatrix} m_{11}U_{N-1}(a) - U_{N-2}(a) & m_{12}U_{N-1}(a) \\ m_{21}U_{N-1}(a) & m_{22}U_{N-1}(a) - U_{N-2}(a) \end{pmatrix}$

where $U_N$ are the *Chebyshev Polynomials* of the second type,

$$U_N(a) = \frac{\sin[(N+1)\cos^{-1} a]}{\sqrt{1-a^2}},$$

and $a = \frac{1}{2}(m_{11} + m_{22}) = \cosh SL \cos kd + \frac{\Delta\beta}{S}\sinh SL \sin kd$. Then:

$$(2.17)\quad M_N = \frac{1}{\sqrt{1-a^2}}\begin{pmatrix} \left[\cosh SL - i\frac{\Delta\beta}{S}\sinh SL\right]e^{ikd}\sin\left(N\cos^{-1}(a)\right) - \sin\left((N-1)\cos^{-1}(a)\right) & -i\frac{\kappa}{S}\sinh SLe^{-ikd}\sin\left(N\cos^{-1}(a)\right) \\ i\frac{\kappa}{S}\sinh SLe^{ikd}\sin\left(N\cos^{-1}(a)\right) & \left[\cosh SL + i\frac{\Delta\beta}{S}\sinh SL\right]e^{-ikd}\sin\left(N\cos^{-1}(a)\right) - \sin\left((N-1)\cos^{-1}(a)\right) \end{pmatrix}$$

Therefore, the transmission coefficient is:

$$(2.18)\quad T_N = [M_N]_{22}^{-1} = \frac{(1-a^2)^{1/2}}{(\cosh SL + i\frac{\Delta\beta}{S}\sinh SL)e^{-ikd}\sin\left(N\cos^{-1}(a)\right) - \sin\left((N-1)\cos^{-1}(a)\right)}$$

For large N, in all cases, the disordered array, ordered array with optimal spacing, and single grating, the long grating regime (*N>>1*) for the transmitted intensity at the centeral wavelength, ($\Delta\beta=0$), is given by:

$$(2.19)\quad \tau_N \propto \exp(-NL_0S) = \tau^N$$

The exponent in the ordered case strongly depends on the grating spacing selection. The strongest reflection is given by the single long grating which has a built in equal spacing zero phase, but anyhow such arrays are difficult to implement. The great surprise is the result for disordered arrays that gives exponential dependence and accordingly high reflectivity, although having its own drawbacks as we describe below.

## 3. The Experiment: System, Measurements and Results

We turn to the experimental study. We first describe the setup for the grating fabrication and then the system for measuring the transmission of the randomly-spaced gratings. The first setup includes a UV laser that is used for grating fabrication, set of lenses that are used to shape the laser beam, mask to create the grating pattern, moving table and controller for fiber placement, and single mode fiber. The second setup is comprised of an Erbium-Doped Fiber Amplifier (EDFA) that is used as a source for the transmission measurement, and an optical spectrum analyzer for conducting the measurements. We then present the measured results that include the transmission spectral behavior and the dependence on the number of gratings. We also show measurements for deducing the one grating transmissivity, needed for the theory verification. At the end of this chapter we discuss the results with a comparison to the localization theory.

### 3.1 The Experimental System: Grating Fabrication and Measurement Setup

The method used to fabricate the gratings is based on a near contact exposure through a phase mask [22]. The setup is illustrated in Fig. 3.1. The UV laser source is an Argon Ion laser of which frequency is doubled by a nonlinear crystal to give a wavelength of 248nm and power of about 200mW. The beam is broadened by a concave lens in order to produce a spot size large enough to illuminate the slit on the mask as uniformly as possible without significantly reducing the beam intensity. Then, the beam is focused in the fiber axis by a cylindrical lens in order to maximize the intensity exposing the fiber. The beam emerging from the cylindrical lens is normally incident on a slit attached to the mask and transfers only 1mm of the beam (the slit is adjusted to pass the interval with the maximum intensity). The slit and mask are placed approximately at the focal point of the cylindrical lens to achieve maximum intensity on the fiber that is adjacent to the mask. The exposing beam is then normally incident on the phase mask and diffracted entirely. The grating is formed by the interference between the +1 and -1 diffracted orders of the phase mask. It was a single mode fiber that was put for a few days inside a high-pressure Hydrogen tank to make it sensitive to photoinduced refractive index change.

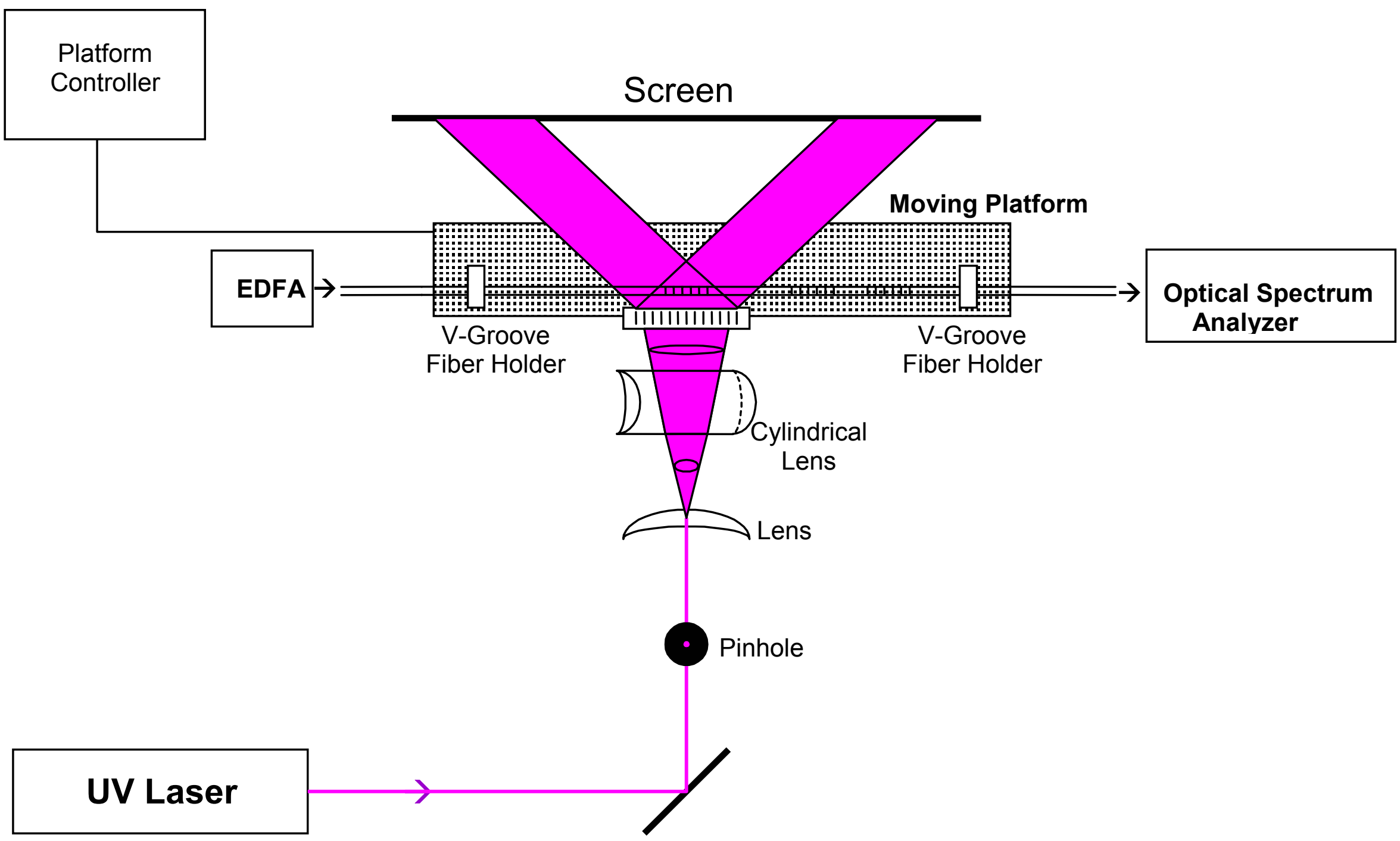

Figure 3.1: The experimental setup: UV laser beam, aligned to a pinhole, is broadened by a lens in order to achieve larger spot size for the grating writing. The beam is then focused at the fiber axes by a cylindrical lens to obtain maximum intensity on the fiber. A phase grating, diffraction gives two 1$^{st}$ order waves, causing a sinusoidal interfering pattern on the fiber. The light transmission measurements in the fiber were done in-situ after each additional grating fabrication. This procedure allowed us to follow the localization buildup with the grating number.

### 3.2 Light Transmission Measurements

In order to achieve reproducibility of the grating spectrum, it was necessary to assure that the exposure time would be similar for all gratings. Therefore, the laser power was adjusted to achieve a relatively slow grating formation time in order to render good accuracy.

The following procedure was repeated for each grating:

- The illumination was started.
- After 1 minute, the illumination was stopped and the spectrum was recorded.
- The automated stage controller was adjusted to move the fiber holding stage a random distance (that was larger than the grating).
- The above procedure was repeated 55 times.

It is noteworthy to mention that the maximum number of 55 gratings was due to limitations of the spectrum analyzer accuracy as the transmitted light intensity decreased with the grating number. The measurement results of the spectrum as well as the intensity are shown in Figs. 3.2-3.5. The experiment was carried out twice for a slightly different exposure time and slightly different grating length (by modifying the distance between the mask and the fiber). The spectra in Figs 3.2 and 3.4 show the detailed wavelength dependence in the grating bandwidth and the gradual loss of the individual Fabry-Perot spectrum along the propagation while acquiring the many and random Fabry-Perot characteristic.

The transmitted power measurements shown in Figs. 3.4 and 3.5 were done at the center of the grating spectrum, where the transmission is minimal, and has been averaged over 0.5*nm* in Experiment no. 1 and over 0.3*nm* in Experiment no. 2. The reason for this averaging stems from the need to overcome the random fluctuations and temperature and stress changes experienced by the fiber during the experiment, causing the measured spectrum to drift and vary. The averaging interval, on one hand, was chosen to be large enough to suppress those environmental changes, but on the other, as the transmissivity magnitude varies with wavelength, the interval had to be limited to a length at which the maximum difference in transmission could be tolerated. In more explicit terms, if the grating minimum transmissivity is ~0.4*dB*, then an accuracy of one magnitude less is tolerable. Furthermore, the noise caused by the optical amplifier and the spectrum analyzer accuracy results in a measured accuracy no better then 0.02$dB$.

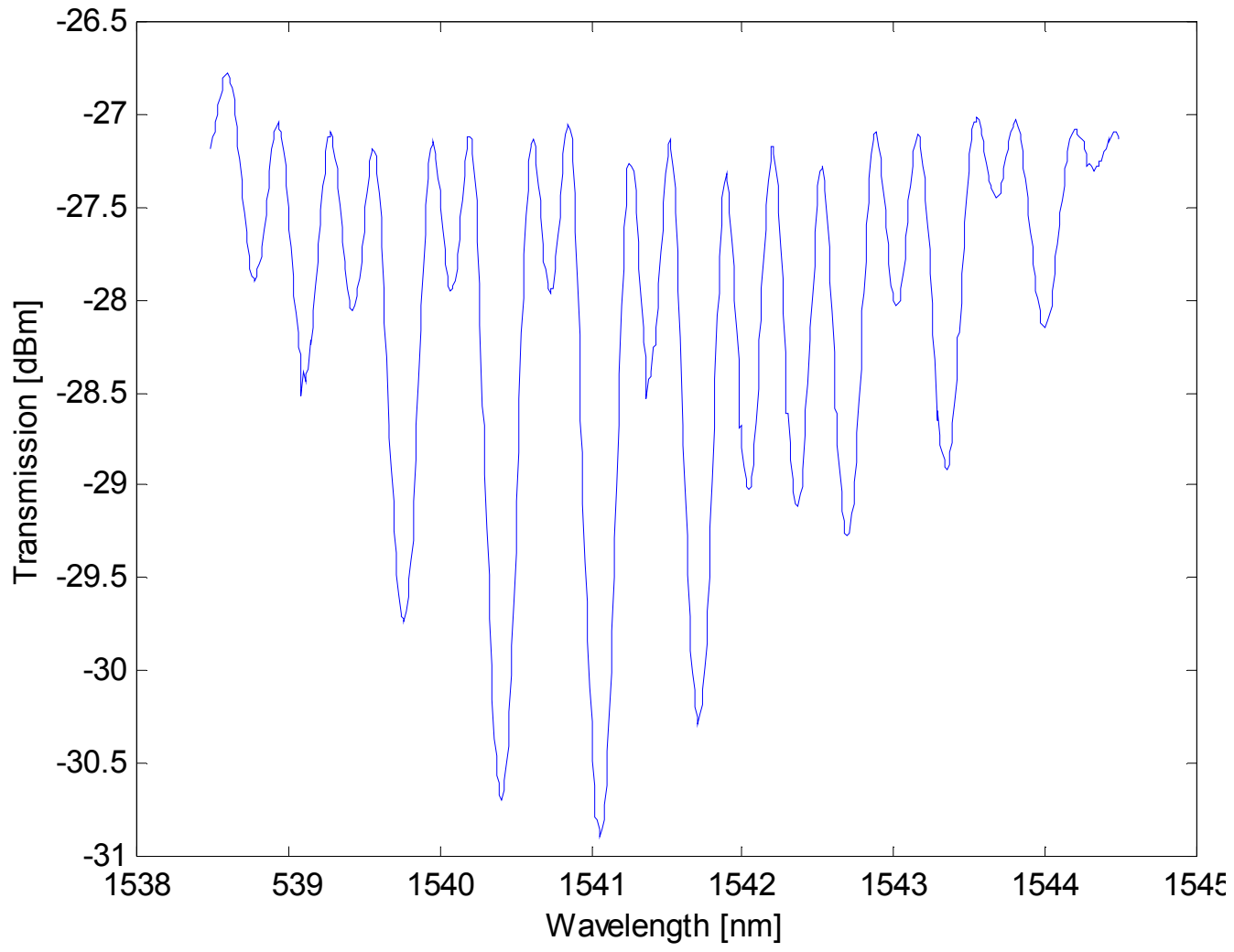

**(a)**

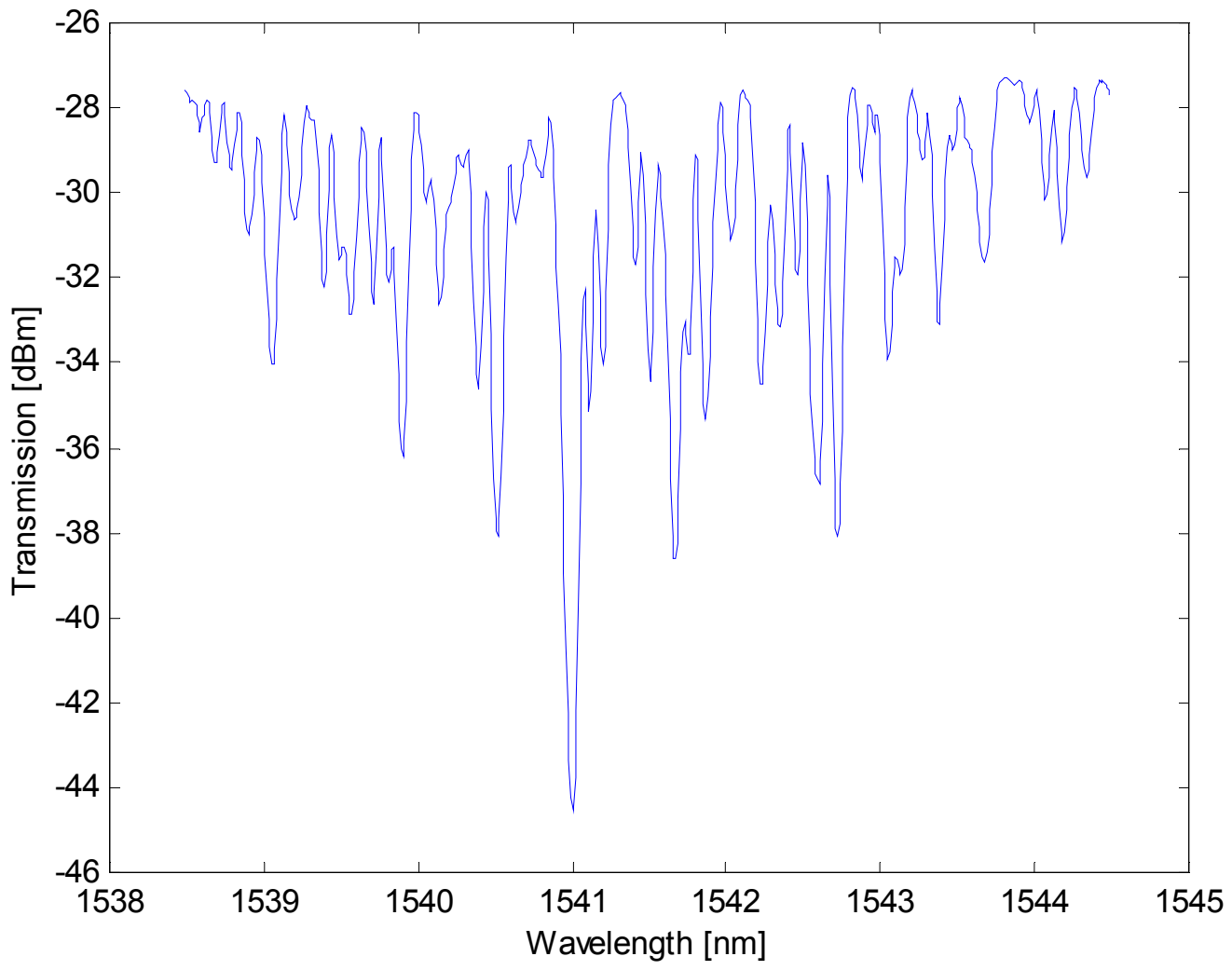

**(b)**

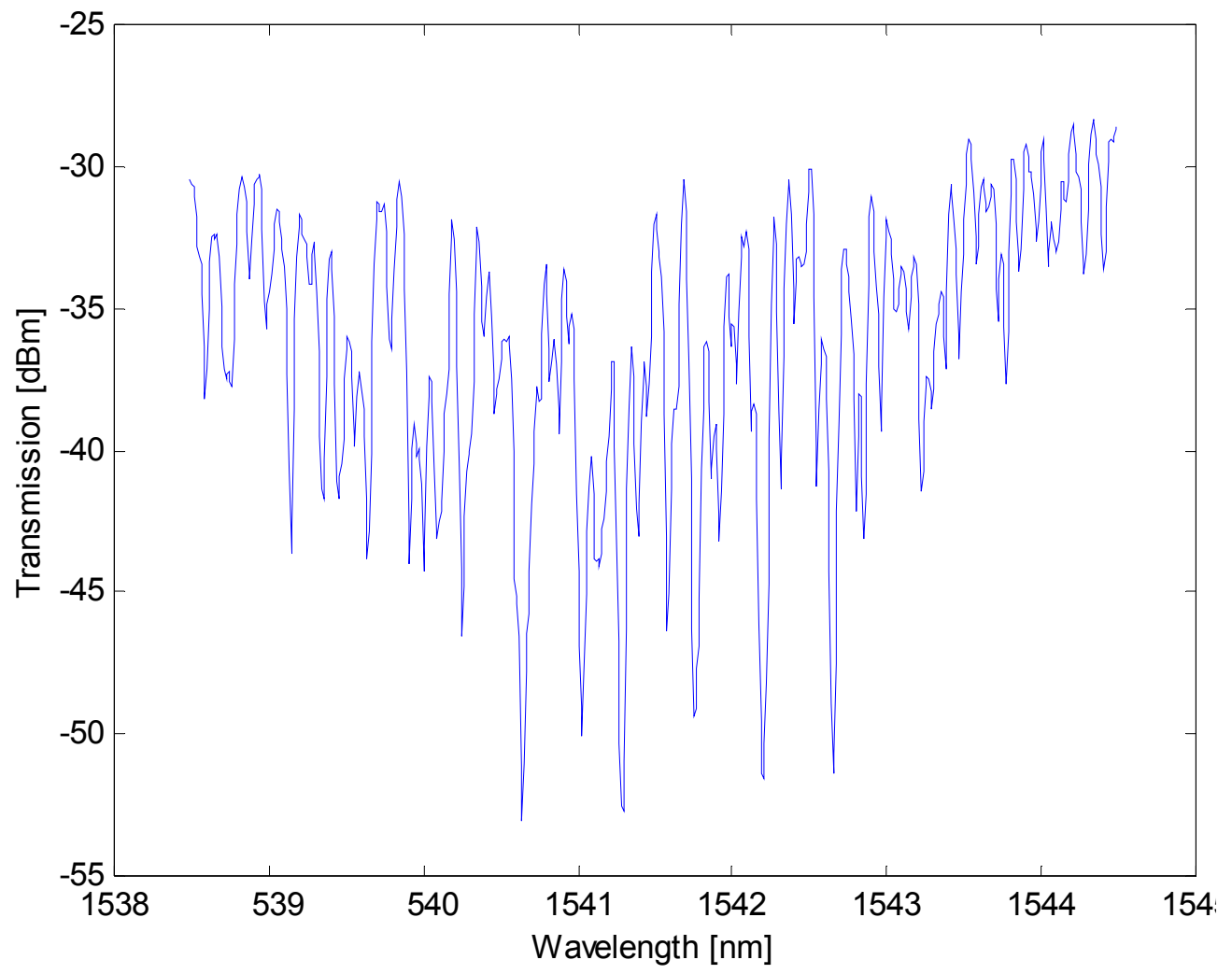

**(c)**

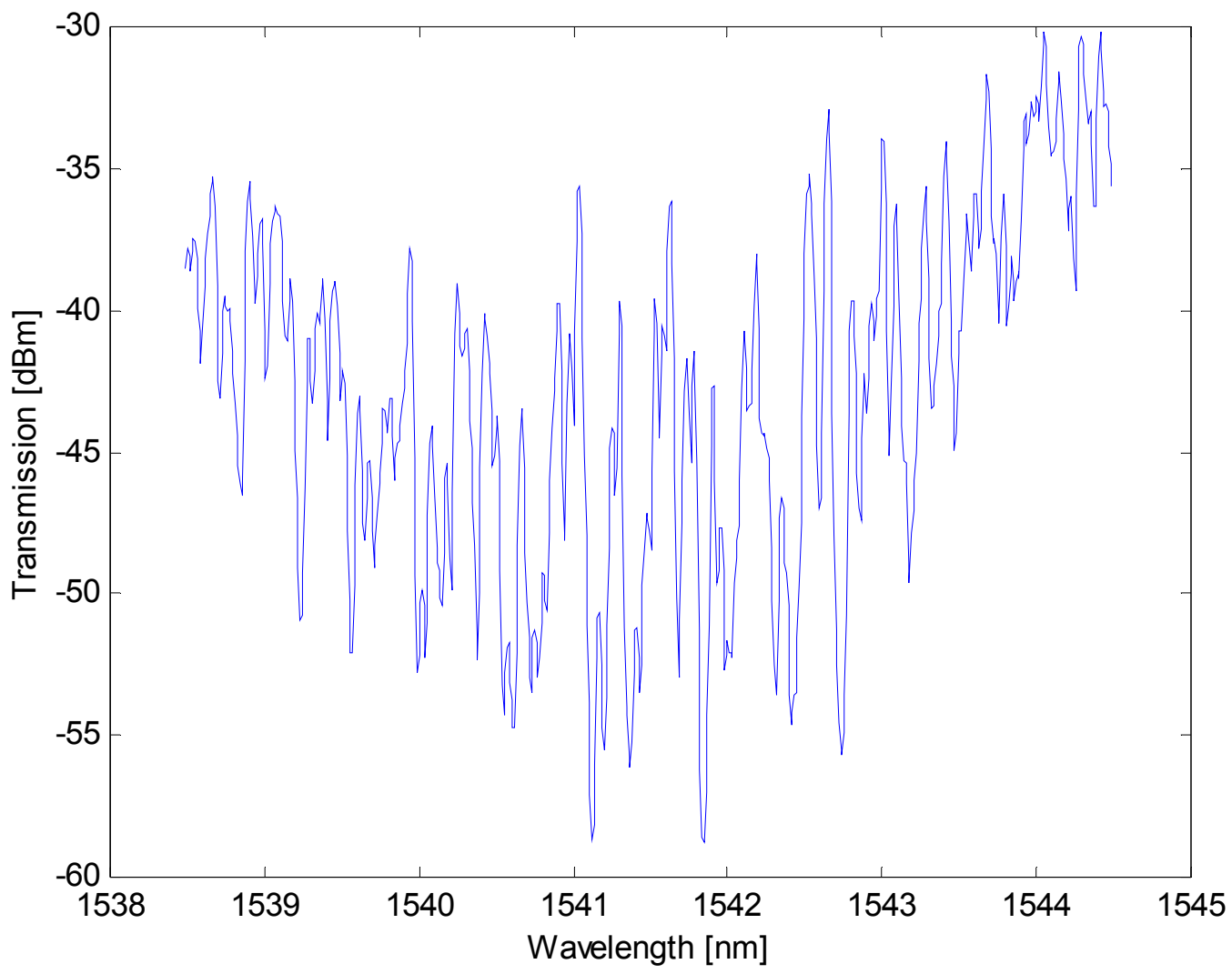

**(d)**

Figure 3.2: Experiment no. 1: Transmitted spectrum measured after (a) 3, (b) 10 , (c) 25 and (d) 50 gratings.

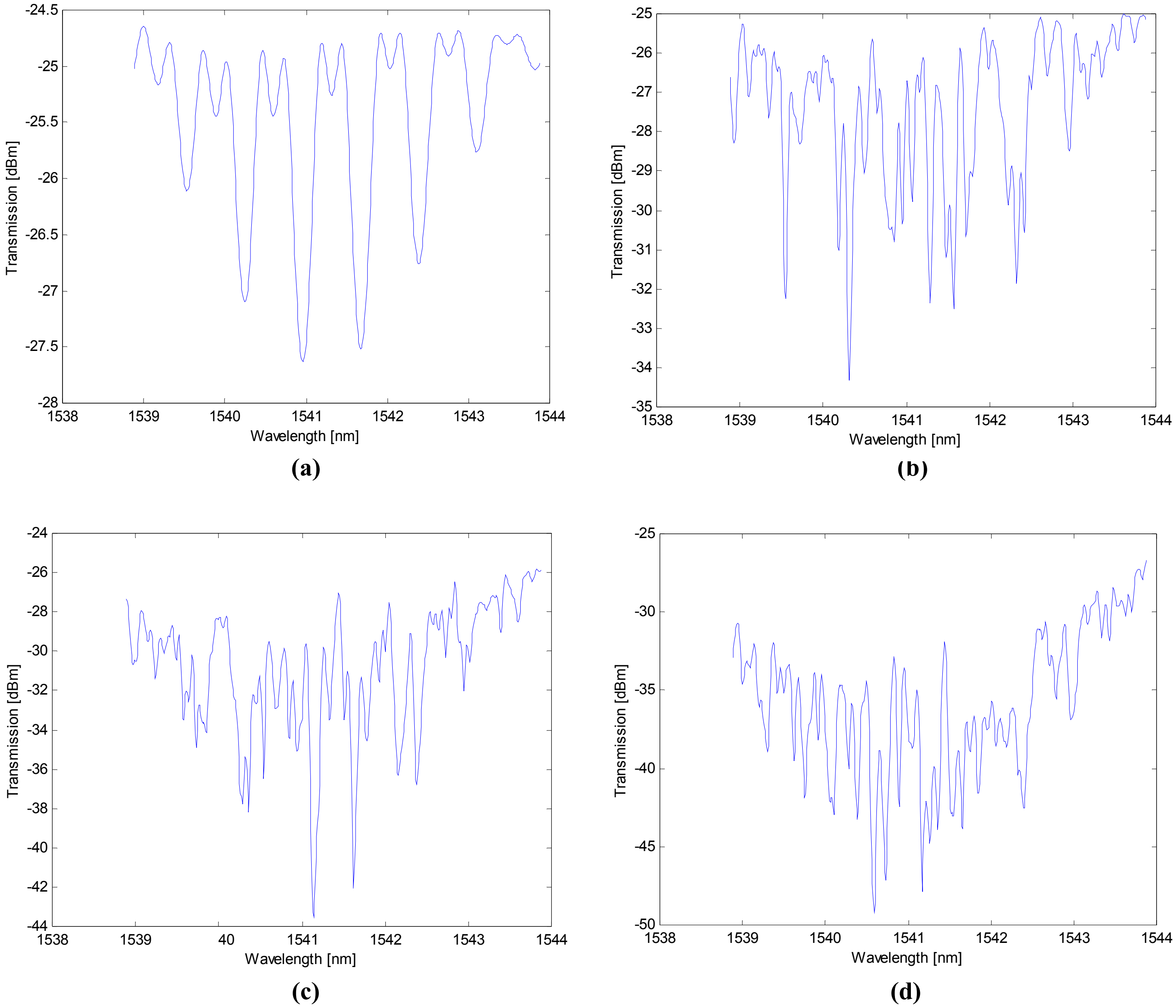

Figure 3.3: Experiment no. 2: Transmitted spectrum measured after (a) 3, (b) 10 , (c) 25 and (d) 50 gratings.

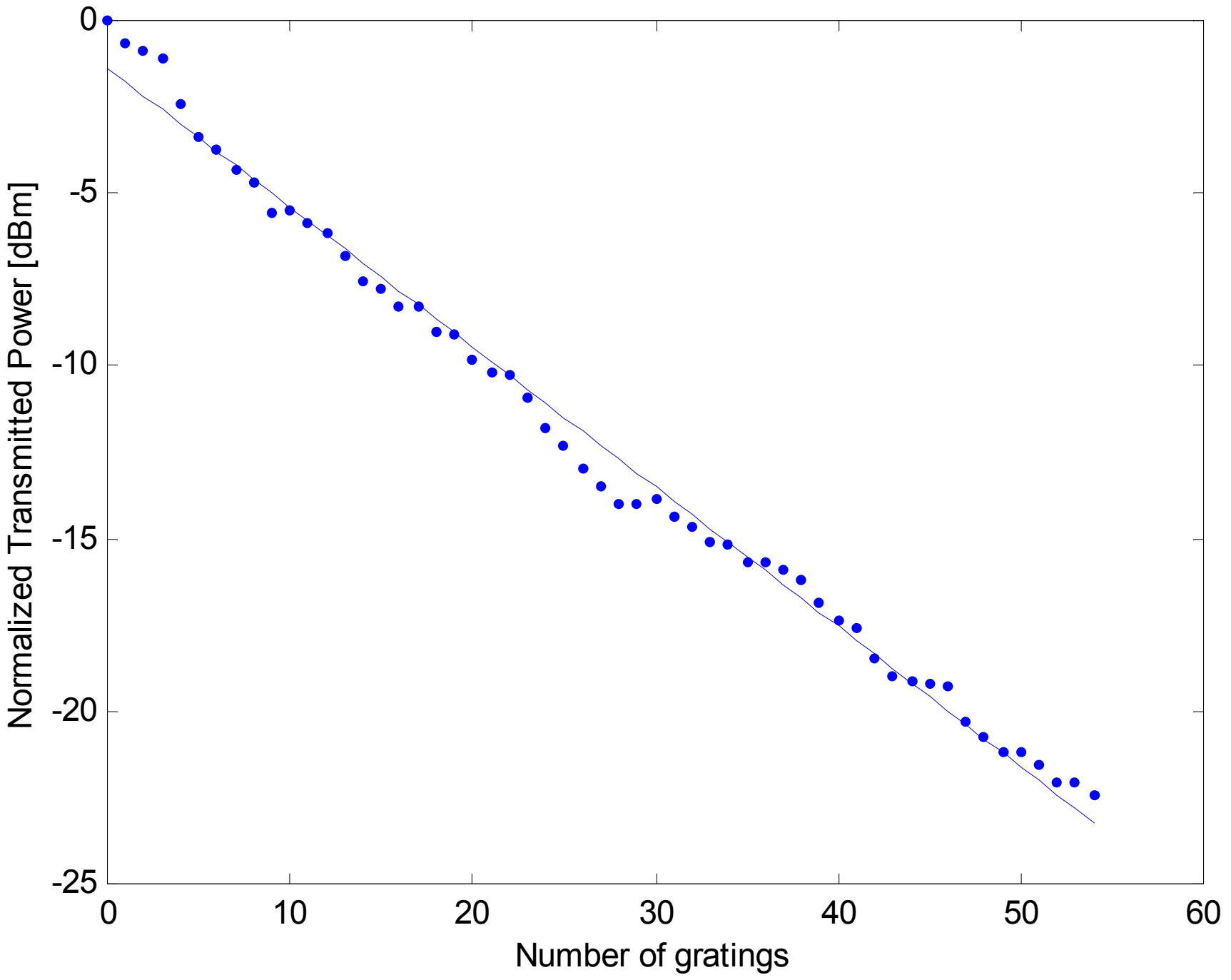

Figure 3.4: Experiment no. 1: The transmission measured at the grating center wavelength (at minimum transmission). The fitted straight line slope is $-0.405$ dB/grating

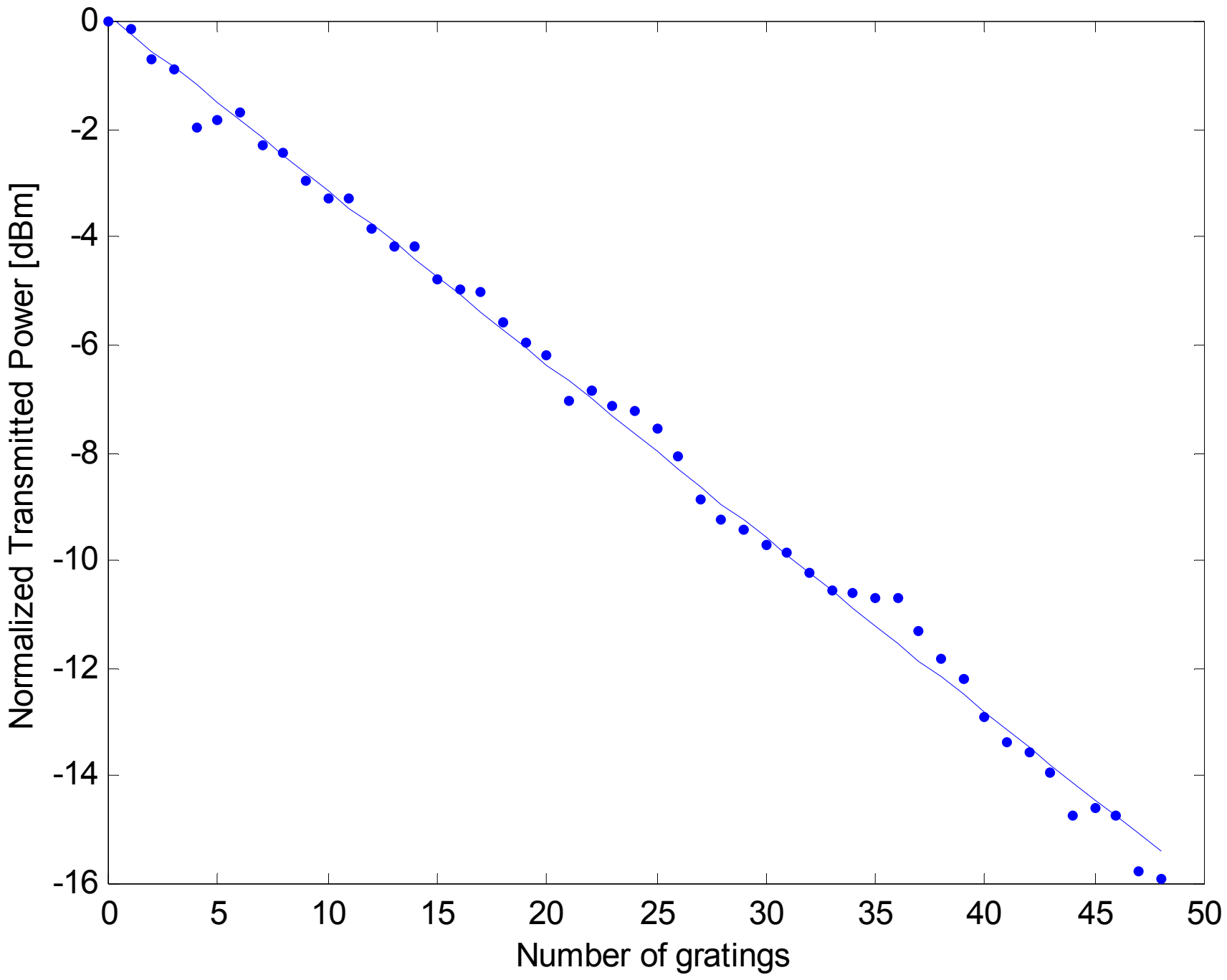

Figure 3.5: Experiment no. 2: The transmission was measured at the grating center wavelength (at minimum transmission). The fitted straight line slope is $-0.326$ dB/grating .

### 3.3 Experiment vs. Theory

Section 2 provided the theoretical asymptotic behavior of the transmission with the exponential decay given by $\tau_N = \exp[-N \ln(1/\tau)]$, where $\tau$ is the intensity transmission coefficient of a single grating. Therefore, in order to compare the theoretical results with the findings obtained from the experiments, it is necessary to first find $\tau$ at the point in the spectrum where the transmission measurements were taken. This single grating value is to be compared to the experimental decay rate of the complete grating array transmissivity.

Measuring the transmissivity of a single grating can be performed by one of two methods. The first and most straightforward method is to take the result obtained from measuring the spectrum of the first grating and normalizing it according to the spectrum measured for a grating-free fiber (that is basically the power spectrum of the EDFA). This method has a great disadvantage as it features wide inaccuracy caused by a power drift that may occur between the two measurements as a consequence of fiber banding, polarization dependent loss, and instability of the source power. While the drift caused by these effects is tolerable for most of the experiment when suppressing it to values of $\sim 0.1 dB$ by maintaining a relatively constant temperature, vibration-free environment, and minimized fiber movement during the experiment, this is not the case when measuring the first grating as the minimum transmission is in a mere tenths of *dB*. The second method is to measure the transmission of two gratings, which is a form of Fabry-Perot, and extracting from it the transmissivity of a single grating. Although this method is less straightforward, it has a great advantage over the previous one as all the effects causing the inaccuracy of the previous method are negligible. This is because the measurement is performed at a specific, given time, without necessitating measurement of the reference level. To achieve good accuracy, the second method was selected. It is now possible to derive the transmissivity of a single grating from the spectrum measurement of two successive gratings. Although the Fabry-Perot properties are simple and known, we derive them here by using the transfer matrix method. The transfer matrix of two successive gratings with a spacing d between them is:

$$(3.1) \qquad M_2 = \begin{pmatrix} \frac{1}{t^*} & -\frac{r^*}{t^*} \\ -\frac{r}{t} & \frac{1}{t} \end{pmatrix} \begin{pmatrix} e^{ikd} & 0 \\ 0 & e^{-ikd} \end{pmatrix} \begin{pmatrix} \frac{1}{t^*} & -\frac{r^*}{t^*} \\ -\frac{r}{t} & \frac{1}{t} \end{pmatrix},$$

where $t_1$ and $r_l$ are the single grating amplitude transmission reflection coefficients, respectively. The distance between the two gratings is *d*, and the propagated field wave number is *k*. Then, the intensity transmissivity of the grating pair can be calculated from $1/T_2 = [M_2]_{22}$, that together with $\tau = |t|^2, \rho = |r|^2$, $\tau = |t|^2$, $\tau_2 = |T_2|^2$ and $\tau + \rho = 1$, $\alpha = kd$, gives the simple Fabry-Perot intensity transmissivity, $\tau_2 = \tau^2 / [\tau^2 + 4\rho \sin^2 \delta]$. Obviously, it is difficult to extract τ from the last expression as it requires knowledge of the transmission coefficient phase and the exact distance between the gratings. However we can easily find it from the minimum of τ as its value doesn't depend on the phase. $\tau_{\min} = \tau^2 / (2 - \tau)^2$. Then the transmission coefficient of a single grating can be written as $\tau = 2\tau_{\min}^{1/2} / (1 + \tau_{\min}^{1/2})$

.

The grating pair measurements of the normalized transmission spectra are shown in Figs. 3.6 and 3.7 for the first two gratings from Experiments 1 and 2, respectively. The value $\tau_{\min}$ is the square root of the minimum transmissivity at the center of the grating spectrum. It is –1.63 *dB* for experiment 1 and –1.41*dB* for experiment 2. Then, for Experiment 1, according to Eq. 3.1, the transmissivity of a single grating is –0.43 *dB*. For Experiment 2 it is –0.35*dB*. According to the localization theory these values are to be compared to the overall transmission slope that are –0.405 *dB*/grating in experiment 1, and –0.326 *dB*/grating for experiment 2.

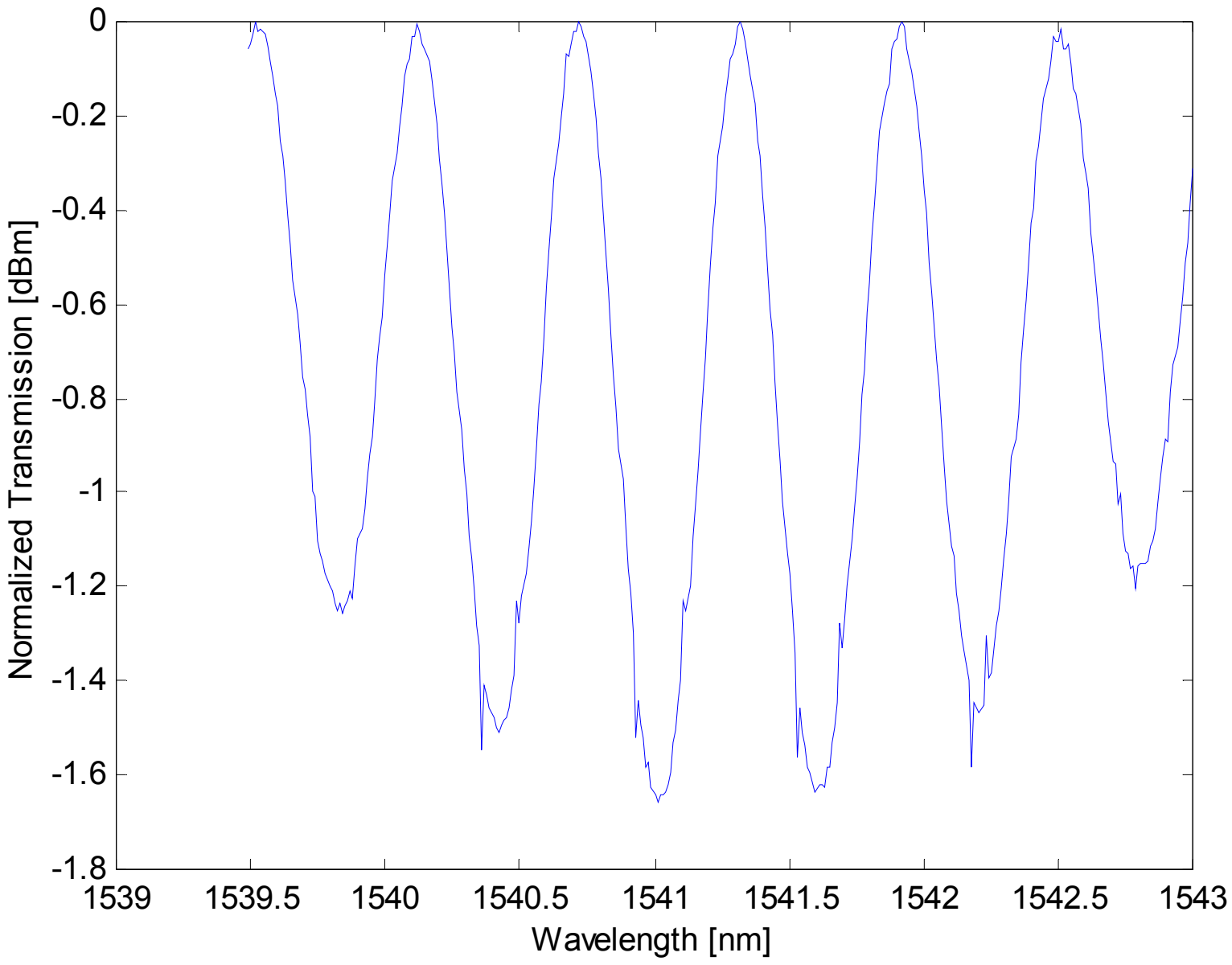

Figure 3.6: Experiment no. 1: The normalized power as measured after two gratings is similar to the spectrum of a Fabry-Perot resonator but with the envelope of the grating spectrum. The transmissivity of a single grating is obtained from the ratio between maximum and minimum transmission power, which in this experiment resulted in $-0.43dB$.

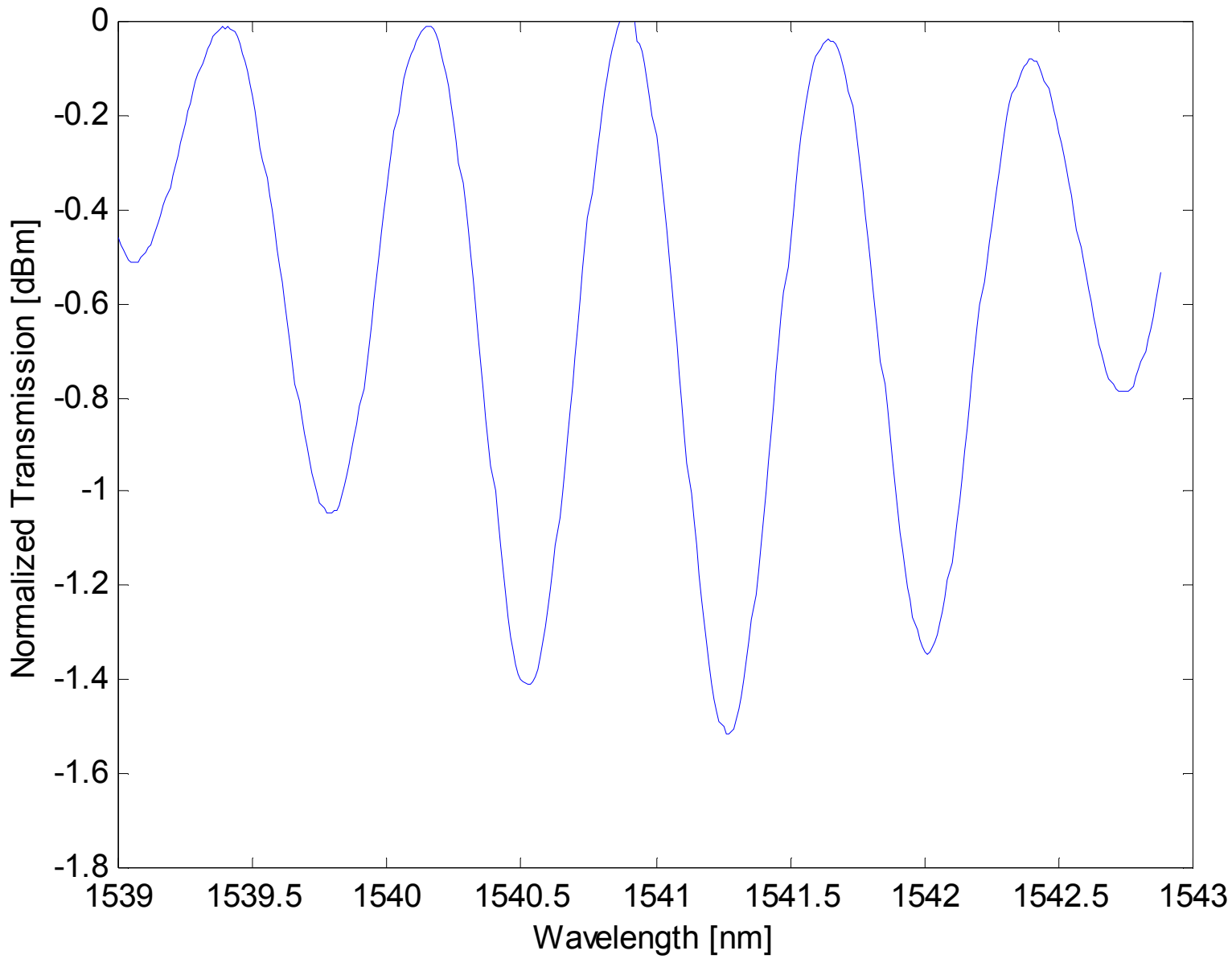

Figure 3.7: Experiment no. 2: The transmissivity of a single grating is obtained from the ratio between maximum and minimum transmission power, which in this experiment resulted in – $0.35dB$.

### 3.4 Loss

Throughout the study, it was assumed that loss is negligible. This assumption should be confirmed experimentally since although the loss of the fiber itself for such short distances is negligible, the process of grating fabrication might introduce some additional losses in the grating formation with the exposure to the UV radiation that changes the fiber uniformity and the absorption coefficient. In order to confirm that the exponential decay indeed resulted from localization, and not from loss, the fiber loss was measured after the grating fabrication. The experiment setup is given in Figure 3.8, where one end of the tested fiber was connected to a spectrum analyzer through a $3dB$ attenuator and 50% coupler, and the other end to the EDFA through a 50% coupler. The intensity measured at the coupler output tap was one-quarter of the reflection from the tested fiber plus one-quarter of the transmission. To evaluate the loss, a measurement was made of the spectrum at the coupler output tap using a regular fiber, and then repeated using the tested fiber. Assuming that a regular fiber can be used as reference for a loss-free fiber, the maximum measured difference between the spectrum of the regular fiber and the tested fiber was less then $0.5dB$ for all wavelengths. As the total transmission measured was $\sim -20dBm$ it may be deduced that the fabricated fiber loss is negligible.

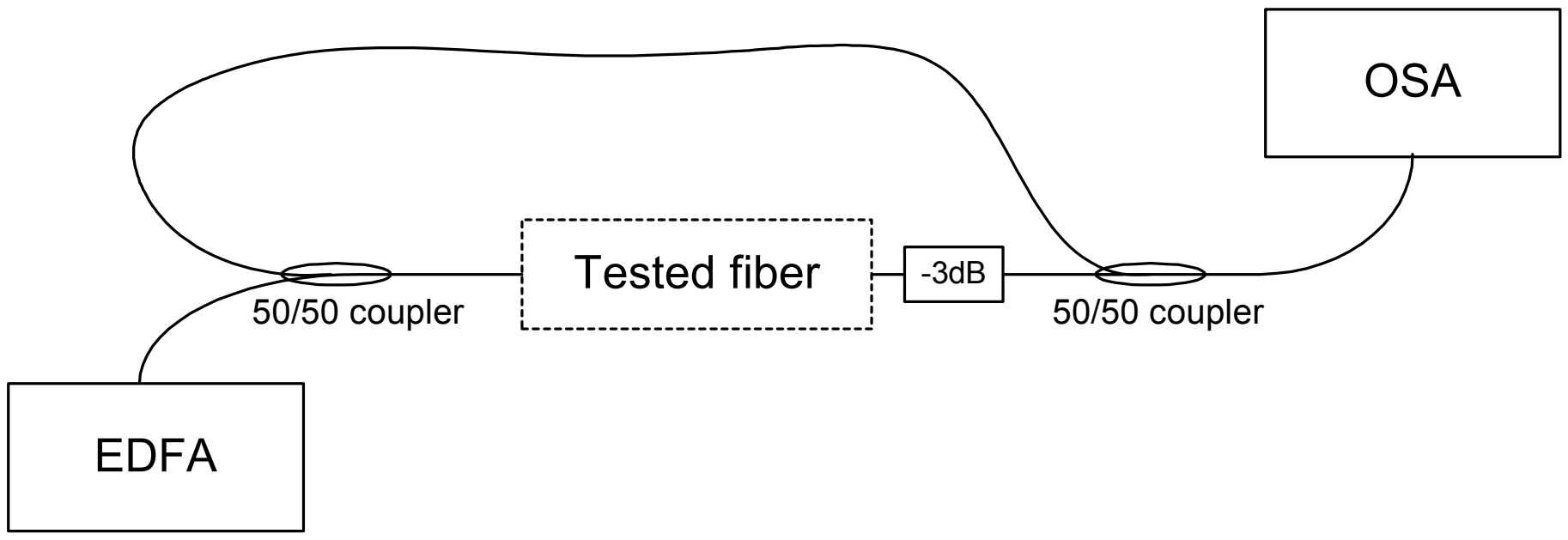

Figure 3.8: Experiment setup for fiber loss measurement: Light reflected from the tested fiber is rerouted back to the left coupler. Half of the reflection is then present at the left tap of the right coupler, and coupled to half of the transmission from the tested fiber. The right tap of the right coupler is connected to the spectrum analyzer and measures one-quarter of the reflection plus the transmission. EDFA is the erbium doped fiber amplifier serves for the light source, OSA is the optical spectrum analyzer.

### 3.5 Experimental Conclusion

The overall results show a good agreement between the experimental results and the theory:

| | Transmissivity of single grating (***dB***) | Total transmissivity slope (***dB*/grating)** |
|---|---|---|
| Experiment no. 1 | **−0.43** | **−0.405** |
| Experiment no. 2 | **−0.35** | **−0.326** |

**Table 3.1: Summary of Experiment Results**

The slight difference between the transmission of a single grating and the total transmission slope results from: (a) The gratings are not exactly equivalent, and therefore the transmission of a single grating (as measured from the first two gratings) can deviate from the average transmission of the gratings. (b) The trasnmission slope is a consequence of a calculation that evaluates the asymptotic behavior of the transmission. For a finite system the transmissivity fluctuates, deviating from the asymptotic calculation. (c) Measurement uncertainties, such as temperature, fiber stress, polarization dependent loss. (d) Stability of the EDFA.

## 4. Summary

We have presented a realization of Anderson localization with light propagating in one-dimensional randomly spaced gratings in a single mode fiber. We described the theoretical analysis and experimentally demonstrated the localization effect. We measured the transmissivity with the exponential decay along the disordered fiber gratings. The magnitude of the decay rate, i.e., inverse localization length, is equal to the log of the inverse single grating transmissivity. The total transmission is comprised only of the wave that passes through all gratings without experiencing any reflections. All other transmitted waves interfere destructively for the transmission. We discussed a ray approach that treats the waves as incoherent due to averaging over random phases in the disordered array, but it fails to adequately describe the special wave interference nature.

We conclude with the application sides of the random grating array. We refer to the reflection side, complementary to the transmission that can become very large with the strong localization effect. Ordered gratings with their filtering and reflection capabilities are widely used in fiber-optics. However it is very difficult to fabricate gratings longer than a few cm. Random grating arrays are by far easier to make with much larger lengths. Here the random nature becomes an advantage. However, even most important feature is that the random array can easily provide very large bandwidth reflection, since it depends on the single grating bandwidth that can be made very short, thus providing very large wavelength bandwidths of tens of nanometers. Another interesting possibility is the use for fiber lasers. The random grating can provide the pseudo cavity for feedback, thus providing a kind of 1D random laser.

**Acknowledgement:**

This work was supported by the Israeli Science Foundation (ISF) of the Israeli Academy of Sciences.

.